\documentclass[printer]{aa} 
\usepackage{lscape}
\usepackage{rotating} 
\usepackage{graphicx}
\usepackage{setspace}
\usepackage{natbib}
\bibpunct{(}{)}{;}{a}{}{,}
\begin{document}

\def\etal{{\it et\thinspace al.}\ }
\def\cm3{cm$^{-3}$\ }
\def\mic {$\mu$m\ }


\title{Radiative
transition rates and collision strengths for
\ion{Si}{ii}
\thanks{Tables 6 and 7 containing the present gf-values, A-values, 
and effective collision strengths
 are only available in electronic form
 at the CDS via anonymous ftp to cdsarc.u-strasbg.fr (130.79.128.5)
 or via http://cdsweb.u-strasbg.fr/cgi-bin/qcat?J/A+A/.
}}
\author{M.A. Bautista\inst{1}
\and P. Quinet\inst{2,3}, P. Palmeri\inst{2}
\and N.R. Badnell\inst{4}, 
J. Dunn\inst{5}, N. Arav\inst{5}}

\institute{Department of Physics, Western Michigan University, Kalamazoo, MI 
49008-5222; manuel.bautista@wmich.edu \and 
Astrophysique et Spectroscopie, Universit\'e de Mons-Hainaut, B-7000 Mons, Belgium
\and 
IPNAS, B15 Sart Tilman, Universit\'e de Li\'ege, B-4000 Li\'ege, Belgium
\and
Department of Physics, University of Strathclyde, Glasgow G4 0NG
\and
Department of Physics, Virginia Polytechnic Institute and State University,
Blacksburg, VA 24061}

 \date{Received date/ accepted date}


\abstract{} 
 {This work reports radiative transition
rates and electron impact excitation collision strengths for levels of the
3s$^2$3p, 3s3p$^2$, 3s$^2$4s, and 3s$^2$3d configurations of \ion{Si}{ii}.}{The radiative data were  computed using
the Thomas-Fermi-Dirac-Amaldi central potential, but with the modifications
introduced by Bautista (2008) that account for the effects of electron-electron
interactions. We also introduce new schemes for the optimization of the variational parameters of the potential.
Additional calculations were carried out with the
Relativistic Hartree-Fock and the multiconfiguration Dirac-Fock methods.
Collision strengths in LS-coupling were calculated in the close coupling approximation
with the R-matrix
method. Then, fine structure collision strengths were obtained
by means of the intermediate-coupling frame transformation (ICFT) method
which accounts for spin-orbit coupling effects.
}{ We present extensive comparisons between the results of different 
approximations and with the most recent calculations and experiment
available in the literature.
From these comparisons we derive a recommended set of $gf$-values and
radiative transition rates with their corresponding estimated uncertainties. 
We also study the effects of different approximations in the representation
of the target ion on the electron-impact collision strengths. Our most
accurate set of collision strengths  
were integrated over
a Maxwellian distribution of electron
energies and the resulting effective collision strengths are given for a
wide range of temperatures. Our results present significant differences from 
recent calculations with the B-spline non-orthogonal R-matrix method. We
discuss the  
sources of the differences.
}
{}
\keywords{atomic data - atomic processes - line: formation -
stars: Eta Carinae - Active Galactic Nuclei}

\titlerunning{Atomic data from the Iron Project LXVI}
\authorrunning{M.A. Bautista et al.}

\maketitle


\section{Introduction}
Singly ionized silicon (\ion{Si}{ii}) is prominent in ultra-violet (UV)
and optical spectra of various astrophysical plasmas.
In terms of absorption spectra, in the spectral range longward of 912 \AA\ 
\ion{Si}{ii} has 8 absorption lines complexes connected to the ground
term 3s$^2$3p $^2$P$^o$. In photoionized plasmas with electron temperatures 
($T_e$)
of the order of $10^4$ K and electron densities ($n_e$) $\le 10^4$ cm$^{-3}$ the
relative optical depths of troughs from the two ground state levels depend
on $n_e$. Hence, this dependence can be used as diagnostic 
of $n_e$ (e.g. Dunn et al. 2009). 
On this regard, the lines
centered at 1814 \AA (3s$^2$3p $^2$P$^o$ - 3s3p$^2$ $^2$D)
are particularly convenient because of
their unusually small oscillator strengths, such that these, among all
other \ion{Si}{ii} lines, are the most likely to be in the linear part of the curve of growth and their column densities can be accurately measured.
Unfortunately, though, the determination of accurate and reliable oscillator
strengths for these transitions is particularly difficult and has been the subject of much 
theoretical and experimental efforts.

\ion{Si}{ii} is also prominent in emission spectra of various kinds of objects.
In the upper chromosphere and lower transition region in the Sun and
late-type stars, line ratios among the \ion{Si}{ii} 1814 \AA\ multiplet
and the intercombination (3s$^2$3p $^2$P$^o$ - 3s3p$^2$ $^4$P) multiplet 
near 2335 \AA\ are potentially useful density diagnostics. However, 
until recently the best electron impact collision strengths of
\cite{dufton} led to predicted line ratios that disagreed with observations
\citep{judge}.
Similarly, theoretical models of emission spectra of Broad Line Regions
of Active Galactic Nuclei fail to reproduce the observed intensities by
factors of a few; a problem that was named by \cite{baldwin} the 
''\ion{Si}{ii} disaster''. 

Despite considerable theoretical work on oscillator strengths there is still considerable spread in 
the results, particularly for those transitions of
most astronomical interest, such as those of the 1814 \AA\ complex. This is because the upper $^2$D 
term of these transitions is made of a mixture of the 3s3p$^2$ and 3s$^2$3d
configurations, which produce strong cancellation 
in the oscillator strengths and makes the f-values very difficult to compute.
Also for these transitions, there has been much spread in experimental gf-values.
Measurements based on the electron beam phase shift method \citep{savage, curtis} could be 
inaccurate for very long lifetimes. 
Absolute emission measurements on an arc \citep{hofmann} are often uncertain owing to the difficult 
control and calibration of the instrument. Determinations
based in comparisons of equivalent widths of the weak 1810 \AA\ lines to  stronger lines in 
astronomical spectra \citep{shull, vanburen} are unreliable 
due to blends and saturation of the stronger lines \citep{jenkins}. The latest, and probably 
most accurate determination of gf-values for these lines was done with the time-resolved 
laser-induced fluorescence technique \cite{bergeson}.

The oscillator strengths for the intercombination transitions (3s$^2$3p $^2$P$^o$ - 3s3p$^2$ $^4$P) have also been subject of controversy. These transitions arise due to spin-orbit mixing of the metastable term 
3s3p$^2$ $^4$P with the $^2$S, $^2$P, and $^2$D terms. The difficulty then comes 
from the fact that the gf-values are sensitive to core-valence and core-core
correlations, in addition to the valence-shell effects. The results of some of 
the most recent calculations \citep{froesefischer} differ from 
experimentally measured transition probabilities \citep{calamai} by $\sim$ 30\%.

But beyond providing accurate oscillator strengths by tailoring atomic structure representations on each type of transition of interest, it is important
to try constructing a single representation that yields reasonably accurate 
energies and oscillator strengths for all levels and transitions considered
simultaneously. This is because various practical applications will
require such a general atomic representation 
for subsequent scattering calculations (e.g. electron impact excitation,
photoionization, and recombination). 
One of the most widely used methods for
this purpose is based on the Thomas-Fermi-Dirac-Amaldi (TFDA) central
potential to generate optimized one-electron orbitals, which represent the 
atomic structure through configuration interaction (CI) expansions and can 
also be used for close-coupling representations of the scattering problem.
Recently, \cite{bautista08} introduced a correction to the TFDA potential
that accounts for the effects of electron-electron interactions
on the radial wavefunction and yield a considerably improved representation of the system. It is thus interesting to apply this new approach to get the
best possible representation of the important \ion{Si}{ii} system.

 In the present paper, to complement and expand the work with the
TFDA potential we also compare with the results of 
the Hartree-Fock Relativistic (HFR), and the multi-configuration Dirac-Fock (MCDF) methods. This general multiplatform approach was successfully employed
in our previous studies of the K-shell spectra of Fe, O, Ne, Mg, Si, S, Ar, Ca, and Ni 
(e.g. Bautista et al. 2003, Garc\'{\i}a et al. 2005, Palmeri et al. 2003a, 2003b, 2008a, 2008b). This has the 
advantage that allows for consistency checks and inter-comparison. It also 
helps to reveal which physical processes are important for any given 
transition, since these different platforms employ different approaches to, for 
example, relativistic effects and orthogonal vs. non-orthogonal orbitals.

The present paper is organized as follows: In the next section we describe the 
calculations of oscillator strengths and transition rates. 
In Section 3 we present the calculations of collision strengths, including some new techniques specifically designed for the present problem. 
Our discussion of the results and conclusions are presented in Section 4. 
 

\section{Radiative Calculations}

The Breit--Pauli Hamiltonian for an $N$-electron system is given
by
\begin{equation}
H_{\rm bp}=H_{\rm nr}+H_{\rm 1b}+H_{\rm 2b}
\end{equation}
where $H_{\rm nr}$ is the usual non-relativistic Hamiltonian, and 
$ H_{\rm 1b}$ and $H_{\rm 2b}$ are the one-body and two-body
operators. The one-body
relativistic operators
\begin{equation}\label{1bt}
H_{\rm 1b}=\sum_{n=1}^N f_n({\rm mass})+f_n({\rm d})+f_n({\rm so})
\end{equation}
represent the spin--orbit interaction, $f_n({\rm so})$, and the non-fine
structure mass-variation, $f_n({\rm mass})$, and one-body Darwin,
$f_n({\rm d})$, corrections. The two-body corrections
\begin{eqnarray}\label{2bt}
\lefteqn{H_{\rm 2b}=\sum_{n>m} g_{nm}({\rm so})+g_{nm}({\rm ss})
                 +g_{nm}({\rm css}) + g_{nm}({\rm d})+}\nonumber \\
                 & & +g_{nm}({\rm oo})\ ,
\end{eqnarray}
usually referred to as the Breit interaction, include, on the one hand, the
fine structure
terms $g_{nm}({\rm so})$ (spin--other-orbit and mutual spin--orbit) and
$g_{nm}({\rm ss})$ (spin--spin); and on the other, the non-fine
structure terms: $g_{nm}({\rm css})$ (spin--spin contact), $g_{nm}({\rm d})$
(Darwin) and $g_{nm}({\rm oo})$ (orbit--orbit).

The oscillator strengths ($f$-values) for dipole allowed transitions have equivalent forms in length and velocity gauges as
\begin{equation}
f_{jk}\equiv{2m(E_k-E_j)\over \hbar^2}|\langle|j|{\bf x}|k\rangle|^2= {i2m\over \hbar^2}
|\langle|j|{\bf v}|k\rangle|^2
\end{equation}
	
The radiative rates ($A$-values) for electric dipole and quadrupole
transitions are respectively given in units of s$^{-1}$ by the expressions
\begin{equation}
A_{\rm E1}(k,i) = 2.6774\times 10^9(E_k-E_i)^3{1\over g_k}S_{\rm E1}(k,i)
\end{equation}
\begin{equation}
A_{\rm E2}(k,i) = 2.6733\times 10^3(E_k-E_i)^5{1\over g_k}S_{\rm E2}(k,i)
\end{equation}
where $S(k,i)$ is the line strength, $g_k$ is the statistical weight of the
upper level, and energies are in Rydberg units and lengths in Bohr radii.

Similarly for magnetic dipole and quadrupole transitions, the $A$-values are
respectively given by
\begin{equation}
A_{\rm M1}(k,i) = 3.5644\times 10^4(E_k-E_i)^3{1\over g_k}S_{\rm M1}(k,i)
\end{equation}
\begin{equation}
A_{\rm M2}(k,i) = 2.3727\times 10^{-2}(E_k-E_i)^5{1\over g_k}S_{\rm M2}(k,i)\ .
\end{equation}
Due to the strong magnetic interactions in this ion, the magnetic dipole
line strength is assumed to take the form
\begin{equation}
S_{\rm M1}(k,i)=|\langle|k|{\bf P}|i\rangle|^2
\end{equation}
where
\begin{equation} \label{m1op}
{\bf P}={\bf P^0}+{\bf P^1}=\sum_{n=1}^N
\{{\bf l}(n)+{\bf\sigma}(n)\}+{\bf P^{\rm rc}}\ .
\end{equation}
${\bf P^0}$ is the usual low-order M1 operator while
${\bf P^{\rm rc}}$ includes the relativistic corrections established
by \citet{dra71}.

In the present work we employ three different computational packages to
study the properties of the strongly correlated configurations 
3s$^2$3p, 3s3p$^2$, 3s$^2$3d, and 3s$^2$4s of \ion{Si}{ii}.

\subsubsection{\sc autostructure}
{\sc autostructure}, an extension by \citet{bad86,bad97} of the atomic
structure program
{\sc superstructure} \citep{eis74}, computes fine-structure
level energies, radiative and Auger rates in a Breit--Pauli
relativistic framework. Single electron orbitals, $P_{nl}(r)$, are
constructed by diagonalizing the non-relativistic Hamiltonian,
$H_{\rm nr}$, within a statistical Thomas--Fermi--Dirac--Amaldi (TFDA) model
potential $V(\lambda_{nl})$ \citep{eis69}. The $\lambda_{nl}$
scaling parameters are optimized variationally by minimizing a
weighted sum of the $LS$ term energies. $LS$ terms are
represented by configuration-interaction (CI) wavefunctions of
the type: 
\begin{equation}
\Psi(LS)=\sum_i c_i \phi_i.
\end{equation}

\cite{bautista08} introduced a modification to the TFDA model potential that accounts in part for the 
effects of electron-electron correlations on the radial wavefunctions by means of additional 
higher order terms in the potential whose strength is controlled by variational parameters. We will refer to this potential as 
c-TFDA.
This formalism also proposed a new optimization technique of the variational parameters by minimizing the difference between predicted, including spin-orbit coupling and relativistic effects, and experimental term averaged energies.
The numerical functional used for this 
optimization was:
\begin{equation}
F = \sum_i {| E_i^{obs}-E_i^{theo} | + \epsilon \over  E_i^{obs}}
\end{equation}\label{functional1}
with $E_i^{obs}$ and $E_i^{theo}$ the observed and theoretical energies  respectively of term $i$ and
$$\epsilon = max\{0,(E_{core}-E_{core}^0)\}  $$
This last term acts as a penalty to the functional whenever the new core
energy exceeds that found from the standard TFDA minimization. This numerical functional was modified 
again for the present work as shown in Section 2.2.


\subsubsection{\sc hfr}
In the Hartree--Fock Relativistic code ({\sc hfr}) by \citet{cow81},
a set of orbitals are obtained
for each electronic configuration by solving the Hartree--Fock
equations for the spherically averaged atom. The equations are the
result of the application of the variational principle to the
configuration average energy. Relativistic corrections are also
included in this set of equations, i.e. the Blume--Watson
spin--orbit, mass-variation and one-body
Darwin terms. The Blume--Watson spin--orbit term comprises the part
of the Breit interaction that can be reduced to a one-body
operator.

The multi-configuration Hamiltonian matrix is constructed and
diagonalized in the $LSJ\pi$ representation within the framework
of the Slater--Condon theory. Each matrix element is a sum of
products of Racah angular coefficients and radial integrals
(Slater and spin--orbit integrals), i.e.
\begin{equation}
\langle a|H|b\rangle = \sum_{i} c_{i}^{a,b} I_{i}^{a,b}.
\end{equation}
The radial parameters, $I_{i}^{a,b}$, can be adjusted to
fit the available experimental energy levels in a
least-squares approach. The eigenvalues and the
eigenstates obtained in this way ({\it ab initio} or
semi-empirically) are used to compute the wavelength and
oscillator strength for each possible transition.


\subsubsection{\sc mcdf}

The multi-configuration Dirac-Fock (MCDF) method considers the Dirac Hamiltonian
for a N-electron atomic system given in a.u. by: 
\begin{equation}
H_D \equiv \sum_{i=1}^N \left[c{\bf \alpha_i} \times {\bf p_i} -(\beta_i-1)c^2-{Z\over r_i}\right ]+\sum_{i\ne j}^N {1\over r_{ij}},
\end{equation}
where c is the speed of light and $\alpha$ and $\beta$ are the Dirac matrices. The atomic state function (ASF) is given as an expansion over jj-coupled configuration state functions (CSFs), which in turn are constructed from Slater determinants built on four-component Dirac orbitals.
The MCDF method is implemented in the computer package {\sc GRASP2K}, described by J\"onsson et al. (2007). 

\subsection{Energy levels and radiative rates}

Our calculations concentrate on the 15 lowest levels, of the
\ion{Si}{ii} system. This is 8 LS terms from configurations 3s$^2$3p, 3s3p$^2$,
3s$^2$4s, 3s$^2$3d, and 3s$^2$4p. A Grotrian diagram of the ions is presented in Fig. 1 which shows the levels and the UV optical and intercombination transitions of interest.

\begin{figure}
\resizebox{\hsize}{!}{\includegraphics{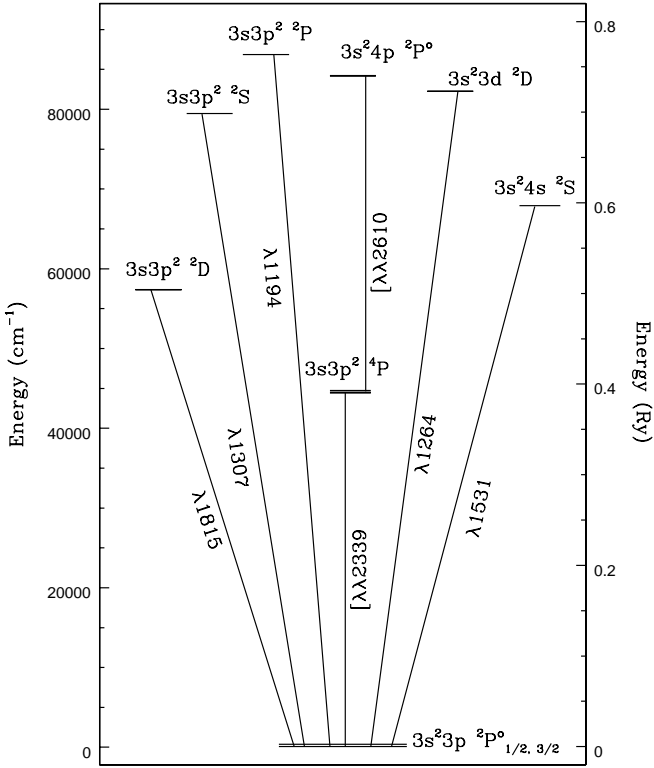}}
\caption{Partial Grotrian diagram of \ion{Si}{ii} showing the 22 lowest
Levels in 8 LS terms and the dipole and intercombination line multiplets 
observable in the UV spectral region.}
\label{fig1}
\end{figure}

\subsubsection{Calculations with HFR} 
     In the physical model adopted in HFR calculations, we suppose that the 
Si$^+$ ion can be represented by three valence electrons surrounding a Ne-like 
Si$^4+$ ionic core with ten electrons occupying the 1s$^2$2s$^2$2p$^6$ closed subshells. 
The intravalence correlation is then considered by the explicit introduction, 
in the model, of the 52 following configurations : 3s23p + 3s24p + 3s25p + 
3s$^2$4f + 3s$^2$5f + 3s3p3d + 3s3p4d + 3s3p5d + 3s3p4s + 3s3p5s + 3s3d4p + 3s3d5p + 
3p$^2$4p + 3p$^2$5p + 3p$^2$4f + 3p$^2$5f + 3d$^2$4p + 3d$^2$5p + 3d$^2$4f + 3d$^2$5f + 3p$^3$ 
(odd parity) and 3s3p$^2$ + 3s$^2$4s + 3s$^2$5s + 3s$^2$3d + 3s$^2$4d + 3s$^2$5d + 3s3p4p + 
3s3p5p + 3s3p4f + 3s3p5f + 3s3d$^2$ + 3s4s$^2$ + 3s4p$^2$ + 3s4d2 + 3s4s5s + 3s3d4s + 
3s3d5s + 3s3d4d + 3s3d5d + 3p$^2$3d + 3p$^2$4d + 3p$^2$5d + 3p$^2$4s + 3p$^2$5s + 3p3d4p + 
3p4s4p + 3d$^2$4s + 3d$^2$5s + 3d$^2$4d + 3d$^2$5d +3d$^3$ (even parity). Core-valence 
correlation is then considered by including a core-polarization (CPOL) 
potential and a correction to the dipole operator as described in many previous 
papers (see e.g. Quinet et al. 1999, 2002). These corrections are used with a 
value of the dipole polarizability equal to 0.16 ai$_0^3$, as computed by Johnson 
et al. 1983 for Si V, and a cut-off radius equal to 0.53 a$_0$ which 
corresponds to the HFR expectation value of $<r>$ for the outermost core orbital, 
i.e. 2p. This method is then combined with a well-established least-squares 
optimization of the radial parameters in order to minimize the discrepancies 
between the Hamiltonian eigenvalues and the available experimental energy 
levels for the 3s23p, 3s$^2$4p, 3s3p$^2$, 3s$^2$4s, 3s$^2$5s, 3s$^2$3d, 3s$^2$4d, 3s$^2$5d and 
3s3p4p configurations. 

\subsubsection{Calculations with MCDF}
Here, we carried out two MCDF calculations. The first one, hereafter referred to as MCDF1, was focused
on our radiative parameters for the intercombination transitions 3s$^2$3p $^2$P$^o$ -- 3s3p$^2$ $^4$P,
and the second one, hereafter referred to as MCDF2, on those of the allowed transitions 3s$^2$3p $^2$P$^o$ -- 3s3p$^2$ $^2$D.

In MCDF1, CI expansions are built including the valence-valence correlation through single and
double excitations from {3s$^2$3p J=1/2,3/2 + 3s3p$^2$ J=1/2,3/2,5/2} extending the orbital active set
up to 5g in 4 steps. The final number of CSFs generated was 2127. For 3 of the  steps, we selected
an "extended optimal level", EOL, optimization option (Dyall et al. 1989 ; Parpa et al. 1996) on the
lowest 5 levels using the same weight for the 5 ASFs. In the first step, no excitation was allowed,
the active set was limited up to 3p, and all the core and valence orbitals were variational. The
second step consisted in extending the expansion through single and double excitations increasing the
active set up to 3d. Here, only the 3d orbital was variational, while freezing all the others to their
values of the preceding optimization step. In the third step, the expansions were further extended increasing the
active set up to 4f. All n=4 orbitals were optimized while freezing all others in a similar
procedure as in the second step. In the final optimization step, only the n=5 were variational freezing all the other orbitals to their preceding step values.

In MCDF2, we used the same procedure as in MCDF1 except that the excitations were from  {3s$^2$3p J=1/2,3/2 + 3s3p$^2$ J=3/2,5/2} and the EOL option was selected on the
first, second, fifth and sixth levels. Here the final number of CSFs used in the MCDF expansions was 1832.

\subsubsection{Calculations with {\sc autostructure}} 

We performed calculations with various different 
configurations expansions, starting with
models similar to those of previously published
work and then evolving to more sophisticated techniques and 
larger configuration expansions. 

\begin{table}[ht]
\centering 
\caption[]{Configurations included in various expansions used
for \ion{Si}{ii}} 
\label{table1}
\begin{tabular}{ll}
\hline\hline
CE1 &
2s$^2$2p$^6$3s$^2$3p, 2s$^2$2p$^6$3s3p$^2$, 2s$^2$2p$^6$3s3p3d,\cr
& 2s$^2$2p$^6$3s3d$^2$, 
 2s$^2$2p$^6$3p$^3$, 
2s$^2$2p$^6$3p$^2$3d, \cr
& 2s$^2$2p$^6$3p3d$^2$, 2s$^2$2p$^6$3d$^3$,
 2s$^2$2p$^6$3s$^2$3d, \cr 
& 2s$^2$2p$^6$3s$^2$4s, 
2s$^2$2p$^6$3s$^2$4p,
2s$^2$2p$^6$3s$^2$4d,\cr
& 2s$^2$2p$^6$3s$^2$5s, 2s$^2$2p$^6$3s$^2$5p,
2s$^2$2p$^6$3s$^2$5d, \cr
&2s$^2$2p$^6$3s3p4s,
2s$^2$2p$^6$3s3p4p,   2s$^2$2p$^6$3s3d4p, \cr
&  2s$^2$2p$^6$3s3d4d,
2s$^2$2p$^6$3s4p$^2$, 
 2s$^2$2p$^6$3s4s5s,\cr
& 2s$^2$2p$^6$3p$^2$4s, 2s$^2$2p$^6$3p3d4p,   2s$^2$2p$^6$3d$^2$4s\cr
\hline
CE2 & 2s$^2$2p$^6$3s$^2$4f,  2s$^2$2p$^6$3s$^2$5f\cr
\hline
CE3 & 2s$^2$2p$^5$3s$^2$3p$^2$, 2s$^2$2p$^5$3s$^2$3p3d,
2s$^2$2p$^5$3s$^2$3p4s\cr
\hline
CE4 &2s$^2$3p$^4$3s$^2$3p$^3$, 2s$^2$2p$^4$3s$^2$3p$^2$3d,
2s$^2$2p$^4$3s$^2$3p$^2$4s \cr
\hline
CE5  & 2s2p$^6$3s$^2$3p$^2$, 2s2p$^6$3s$^2$3p3d, 2s2p$^6$3s$^2$3p4s \cr
\hline
CE6 &2s$^2$2p$^5$3s$^2$3p4p, 2s$^2$2p$^5$3s$^2$3p4d,
2s$^2$2p$^4$3s$^2$3p$^2$4p, \cr 
& 2s$^2$2p$^4$3s$^2$3p$^2$4d, 
2s$^2$2p$^3$3s$^2$3p$^4$, 2s$^2$2p$^3$3s$^2$3p$^3$3d, \cr 
& 2s$^2$2p$^3$3s$^2$3p$^3$4s, 2s$^2$2p$^3$3s$^2$3p$^3$4p,
 2s$^2$2p$^3$3s$^2$3p$^3$4d \cr
\hline
\end{tabular}
\end{table}

Table 1 presents the configurations included in 7 basic expansions
studied. The various configuration expansions (CEs) are named CE1, CE2, ..., CE6
and are tabulated in such a way that each CE includes the configurations
of all previous CEs. In other words, CE1 includes the 24 configurations
listed at the top of Table 1, CE2 includes all 24 configurations from CE1
plus 2s$^2$2p$^6$3s$^2$4f, CE3 includes the 25 configuration from CE1 and CE2 plus 3 more, and CE6 includes all 48 configurations in the table.

\begin{table*}[Ht]
\centering 
\caption[]{Comparison of level energies (Ry) for \ion{Si}{ii}} 
\label{table2}
\begin{tabular}{lrrrrrrrrrr}
\hline\hline
Level &Exp.    & AST1& AST2& AST3& AST4& AST5& AST6& AST7& AST8& AST9\\ 
      &        & \multicolumn{9}{c}{Difference from experiment(\%)}\\ \hline
3s$^2$3p~$^2$P$^o_{1/2}$& 0.0 & 0.0 & 0.0 & 0.0 & 0.0 & 0.0 & 0.0 & 0.0 & 0.0 & 0.0 \cr
3s$^2$3p~$^2$P$^o_{3/2}$&0.002618 &-18.8 &-17.8 &-14.8 &-15.0 &-15.0 &-15.0 &-15.1 & 0.19 &-0.12 \cr
3s3p$^2$~$^4$P$_{1/2}$  &0.390244 &-7.86 &-17.4 &-11.3 &-11.3 &-10.9 &-10.9 &-10.6 &-7.06 &-7.62 \cr
3s3p$^2$~$^4$P$_{3/2}$  &0.391231 &-7.90 &-17.4 &-11.4 &-11.3 &-10.9 &-10.9 &-10.6 &-7.01 &-7.57 \cr
3s3p$^2$~$^4$P$_{5/2}$  &0.392828 &-7.96 &-17.4 &-11.4 &-11.3 &-10.9 &-10.9 &-10.6 &-6.92 &-7.48 \cr
3s3p$^2$~$^2$D$_{3/2}$  &0.504016 & 14.5 & 2.42 & 1.03 & 1.00 & 1.18 & 1.18 & 1.42 &-0.01 &-0.25 \cr
3s3p$^2$~$^2$D$_{5/2}$  &0.504160 & 14.4 & 2.40 & 1.01 & 0.98 & 1.15 & 1.16 & 1.39 &-0.02 &-0.27 \cr
3s$^2$4s~$^2$S$_{1/2}$  &0.596884 &-0.53 & 3.31 &-0.39 &-0.49 &-0.59 &-0.54 &-0.32 & 15.3 &-4.64 \cr   
3s3p$^2$~$^2$S$_{1/2}$  &0.698626 & 0.69 & 1.64 & 3.23 & 3.25 & 3.46 & 3.46 & 3.60 & 2.73 & 1.78 \cr
3s$^2$3d~$^2$D$_{3/2}$  &0.722986 & 35.3 &-2.37 & 1.19 & 1.13 & 1.14 & 1.17 & 1.29 & 11.1 & 10.3 \cr   3s$^2$3d~$^2$D$_{5/2}$  &0.723136 & 35.2 &-2.39 & 1.17 & 1.12 & 1.12 & 1.16 & 1.28 & 11.1 & 10.3 \cr
3s$^2$4p~$^2$P$^o_{1/2}$&0.739870 & 22.0 & 3.40 & 6.77 & 6.83 & 7.03 & 7.07 & 6.81 & 13.9 & 13.7 \cr
3s$^2$4p~$^2$P$^o_{3/2}$&0.740416 & 22.3 & 3.41 & 6.82 & 6.87 & 7.08 & 7.12 & 6.86 & 14.0 & 13.9 \cr
3s3p$^2$~$^2$P$_{1/2}$  &0.763660 & 0.75 & 5.74 & 5.52 & 5.49 & 5.68 & 5.71 & 5.74 & 4.20 & 5.01 \cr
3s3p$^2$~$^2$P$_{1/2}$  &0.765503 & 0.69 & 5.68 & 5.47 & 5.44 & 5.64 & 5.66 & 5.70 & 4.21 & 5.04 \cr
\hline
\end{tabular}
\end{table*}

\begin{figure}
\resizebox{\hsize}{!}{\includegraphics{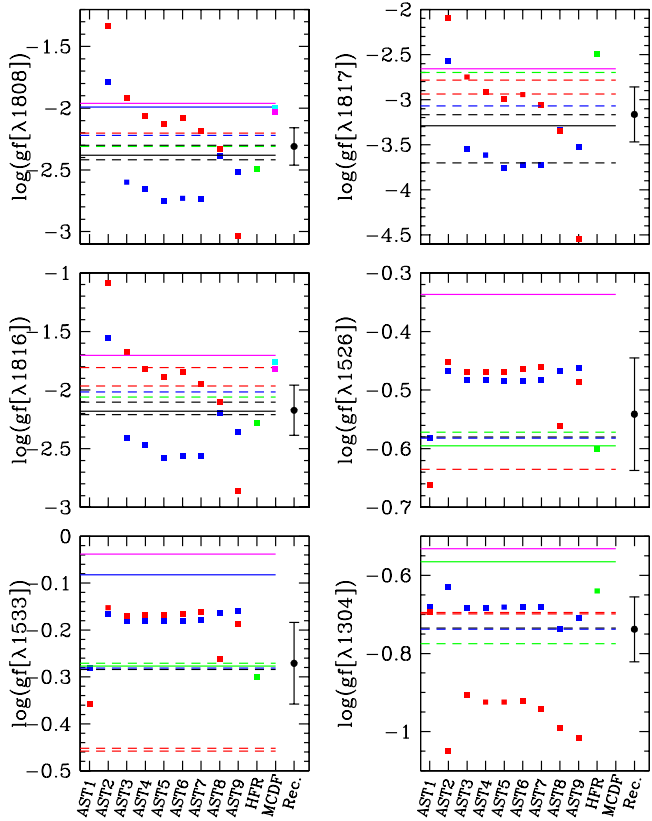}}
\caption{Evolution of the $gf$-values for transitions in
\ion{Si}{ii}
from our different calculations AST1, ..., AST9 and comparison with previous determinations. The square points depict our results in the length gauge
(blue) and velocity gauge (red). The solid horizontal lines
depict experimental determinations by \cite{shull}
(magenta), \cite{vanburen} (blue), \cite{schectman} (green), \cite{bergeson} (black). The dashed lines depict theoretical results of \cite{froesefischer},
(blue), \cite{tayal07} (red), \cite{nahar} (green).}
\label{gdipole}
\end{figure}

\begin{figure}
\resizebox{\hsize}{!}{\includegraphics{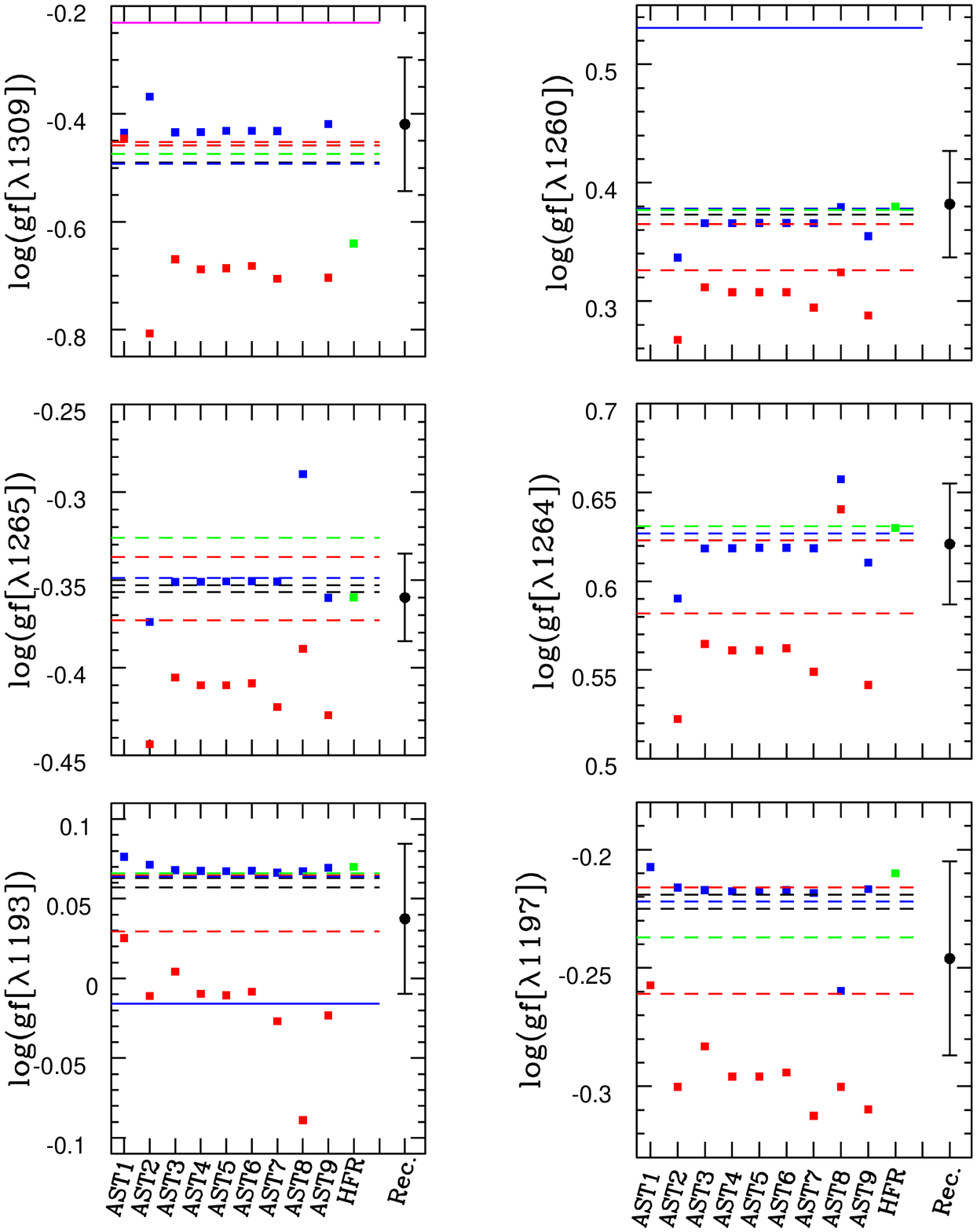}}
\caption{Cont. Fig. \ref{gdipole}}
\label{gdipole2}
\end{figure}

The rational for these configurations and the progression of various calculations carried is as follows:

\noindent{-} {\bf AST1}: This calculation uses the expansion CE1 that keeps the 2s
and 2p shells closed and promotes the three remaining electrons among orbitals
with principal quantum number $n$=3, 4, and 5 and orbital angular momentum
$l$=0, 1, 2. The TFDA potential was used and the orbitals were optimized in the
standard {\sc autostructure} procedure minimizing the energies of the lowest 8 LS terms. Fine structure coupling and relativistic corrections are introduced
as perturbations after optimizing the orbitals. From this calculation the core energy of the system is -578.40 Ry. 
Table~1 compares the calculated energies relative to the ground level with experimental values
from NIST (2008). The results are rather unsatisfactory with energy differences 
scattered between -19\% and 35\%. Even more troublesome are the $gf$-values
that result from this calculation, particularly those for the 1814\AA\ multiplet, which are  
considerably overestimated with respect to the experimental
determination of \cite{bergeson} (see Fig. 2). This is because it is difficult to represent
3s$^2$3p and 3s3p$^2$ states simultaneously with orthogonal orbitals, owing to 
polarization and orbital relaxation effects. 

\noindent{-} {\bf AST2}: For this calculation we use the same expansion CE1 as in
AST1 but with the use of the c-TFDA potential of \cite{bautista08} optimized
against the experimental term energies using the functional of Eqn. \ref{functional1}. This gave a core energy of -578.34 Ry, which is slightly higher than in AST1. Yet, the calculated energies here agree much better with experiment than from AST1, with most terms within 5\% of experiment with the
exception of the two excited levels 3s$^2$3p $^2$P$^o_{3/2}$, from the ground
multiplet, and 3s3p$^2$ $^4$P$_j$. 
Together with the changes in calculated energies there are also significant changes in the predicted 
$gf$-values for the 1814\AA\ and 1531\AA\ multiplets, as illustrated in 
Fig. 2. 

\noindent{-} {\bf AST3}: This calculation is just like AST2, but we now add the
configurations 3s$^2$4f and 3s$^2$5f (i.e. CE2). We find that the f orbitals have an important effect on increasing the polarizability of configurations that include d orbitals, such as the 3s3d$^2$, and 3p$^2$3d. Thus, by including $n$f configurations in the expansion we find different optimization for the core. When including these $n$f orbitals one has to include enough $n$d configurations in order to reach convergence in the solution. But, various configurations such
as 3d, 3d2, 4d, and 5d are already included since CE1. In the course of the calculations we checked on the contributions of 5d and 5f and they were found to be small. Thus, we believe that the calculation is nearly converged. In this calculation
the core energy improves to -578.38 Ry and the agreement between calculated 
and experimental energies with respect to the ground level also improves a little. Despite the 
relatively minor changes in the calculated energies between this calculation and the previous one there are 
large changes in the $gf$-values for the 1814\AA\ and 1264\AA\ multiplets, with a
drop of $\sim$0.8 dex for the former and a rise of $\sim$0.9 dex for the latter.
 
\noindent{-} {\bf AST4}, {\bf AST5}, {\bf AST6}, and {\bf AST7}: For these calculations we progressively 
add new configurations by opening the 2s and 2p orbitals, with one 2p electron promotions in AST4 (CE3), 
two 2p electron promotions in AST5 (CE4), one 1s electron promotions in AST6 (CE5) and three 2p 
electron promotions in AST7 (CE6). By this point the
computational time in optimizing all orbitals of the potential has become
exceedingly long, going
from a few minutes in the AST1 calculation to several hours in the AST7 calculation. For this reason, and with the purpose to see the effects
of increasing CI we did not optimize again the orbitals for these calculation, but instead chose the same scaling parameters as in AST3. Despite the large increase in CI in these computations we find 
no significant change in either the calculated energies, see Table 2. 
The calculated core energies also remain constant at -578.38 Ry for
{\bf AST4}, {\bf AST5}, and {\bf AST6} and -578.39 Ry for {\bf AST7}. 
The computed $f$-values are also unaltered by CI, except
for those of the 1814\AA\ multiplet which seem to converge asymptotically.
Nonetheless, for this and most of the other transitions the difference between
length and velocity gauges of the $f$-values is uncomfortably large.  
It is clear now that increasing the amount of
CI in the calculations beyond that of the AST7 calculation will not improve 
the quality of the results. Instead, we need to turn our attention towards the
optimization of the orbitals.

\noindent{-} {\bf AST8}. For this calculation we decided to return to the simplest expansion CE1, but 
try a different optimization technique. Because the $f$-values for the 1814\AA\ multiplet is so sensitive to cancellation in the mixing of the 3s3p$^2$ $^2$D and 3s$^2$3d $^2$D states it 
is crucial to reproduce the experimental energies of the levels as accurately
as possible. Thus, we modified the {\sc autostructure} code use an 
optimization functional of the form
\begin{equation}
F = \sum_i {w_i(E_i^{obs}-E_i^{theo})^2 + \over  (E_i^{obs})^2}+\epsilon,
\end{equation}\label{functional2}
where the sum goes over fine structure energy levels and $w_i$ is a user defined 
weight on different levels. For our calculation we ran this sum for the lowest
15 levels of the \ion{Si}{ii} system and we weighted the 3s$^2$3p $^2$P$^o_{3/2}$
and 3s3p$^2$ $^2$D$_J$ levels 5 times as much as the rest. This kind of optimization provides much better energies for all levels and nearly exact energies for the 3s$^2$3p $^2$P$^o_{3/2}$ and 3s3p$^2$ $^2$D$_J$ levels (within
0.2\% of experiment). This optimization also provides a lower core energy than
from previous calculations (-578.42 Ry). Nevertheless, this optimization does not reduce the difference between length and
velocity forms of the $f$-values.

An alternative optimization functional that is already available in {\sc autostructure} is based on the differences between length and velocity 
$f$-values. This, however, does not work because it leaves the level energies 
to change way off from experiment. 

Then, we define a combined functional of the form
\begin{eqnarray}
\lefteqn{F= \sum_i {w_i(E_i^{obs}-E_i^{theo})^2 + \epsilon \over  
(E_i^{obs})^2}}\nonumber \\
 & & + \sum_i\sum_j {w_i(gf_{ij}^{l}-gf_{ij}^{v})^2\over
(gf_{ij}^{l}+gf_{ij}^{v})^2},
\end{eqnarray}\label{func2}
where $gf_{ij}^{l}$ and $ gf_{ij}^{v}$ are the length and velocity forms of the
$gf$-value for the transition $i\to j$.
An optimization of this kind heavily weighted on the $gf$-values for the
1260\AA\ multiplet yields length and velocity values in very close agreement 
with each other. These results also agree very well with the experimental determination
of \cite{bergeson}. 
Unfortunately, the obtained $gf$-values for transitions involving higher terms
seem to deteriorate. This is to be expected, as the small expansion used here misses
important core-valence and core-core correlations.
 
\noindent{-} {\bf AST9}: For this last calculation we use the largest expansion CE6 and
optimize the orbitals as in {\bf AST8}. For a model expansion of this size
numerical optimization becomes challenging, and the code took 4 days to do
so. The resulting energies are slightly better than in the previous calculation,
and the core energy is predicted at -572.41 Ry. It is
found that the $f$-values in the length gauge for the 1814\AA\  multiplet remain
in good agreement with the experiment of \cite{bergeson} although the velocity
$f$-value departed somewhat. Interestingly, for these transitions the present velocity $f$-value
is lower than the length $f$-value, contrary to what was seen in previous calculations. The present difference between the two forms of the $f$-value
probably owes to the need of finer optimization of the orbitals, which became
computationally prohibitive for the large CI expansion used here. Another
interesting set of $f$-values are those for the 1260\AA\ multiplet, which in the present
calculation result somewhat higher and with greater difference between length and velocity than from previous calculations. The present result seems to
favor the experimental determination of \cite{schectman}, while the results of previous calculations seemed very stable and in good agreement with
\cite{froesefischer} and \cite{tayal07}.

\

\begin{figure}
\resizebox{\hsize}{!}{\includegraphics{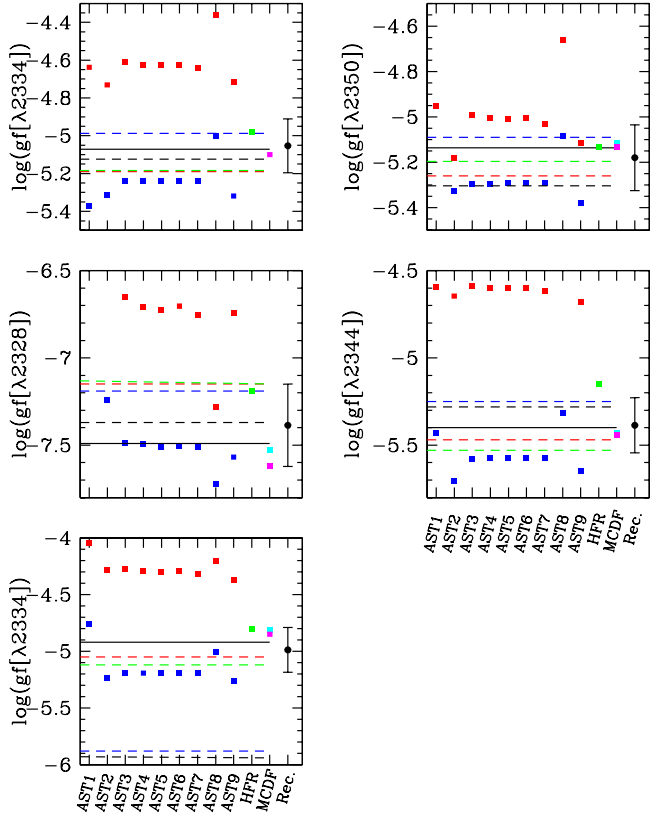}}
\caption{Evolution of the gf-values for intercombination transition in
\ion{Si}{ii}
from our different calculations AST1, ..., AST9 and comparison with experimental values of \cite{calamai}. The square points depict our results in the length
gauge (blue) and velocity gauge (red). The experimental vales are given by the
horizontal solid line, with quoted errors given by dashed lines.}
\label{intercomb}
\end{figure}

At this point is worth pointing out the detailed study of \cite{dufton92} on the LS f-value for
the 1814~\AA\ transition. They found that in dealing with the cancelation effects from contributions from the 3s3p$^2$~$^2$D and 3s$^2$3d~$^2$D states one must pay attention to the calculated energy difference between these states. In this sense we notice that our prefer calculation AST8 and AST9 yield energy separations for the 3s3p$^2$~$^2$D and 3s$^2$3d~$^2$D states that differ by $\sim10-11$\% from experiment, in contrast
with calculations AST3 through AST7 that reproduce these energy separation significantly better. However, getting exact energies is neither sufficient nor guarantee of resulting accurate gf-vaues. Inspection of Table 1 of Dufton et al. (1992) shows that essentially the same gf-value can obtained when the energy difference with respect to experiment was less than 2\% and nearly 11\%. By contrast, calculations that predicted energies differing from experiment by 5 and 7\% got gf-values that were off by roughly a factor of 3. Moreover, we are mostly concerned with the fine structure gf-values and for that we also need accurate energies for the J resolved levels of the 3s$^2$3p~$^2$P and 3s3p$^2$~$^2$D terms and these are best reproduced by the AST8 and AST9 calculations, as opposed to the AST7 calculation that underestimates the energy of the 3s$^2$3p~$^2$P$_{3/2}$ level by 15\%.

On the other hand, it is also known that in computing gf-values one may not rely only
on the agreement between length and velocity values, because such an criterium can be misleading if the model potential is oversimplified. Nonetheless,
we believe that the agreement between length and velocity results in our calculations
AST8 and, particularly AST9 is very significant. AST9 uses the largest CI expansion that we could handle, which the inner core orbitals to account for polarization. In addition, this calculation uses the modified TFDA potential to account for additional electron-electron correlation. Thus AST9 employs the most complex and complete potential that we can build at this time, then the good agreement between length and velocity seems meaningful.

Now we look at the results for the intercombination transitions 3s$^2$3p $^2$P$^o_j$ -- 3s3p$^2$ $^4$P$_{j'}$. These lines, seen in emission in the 2328 -- 2350\AA\ spectral range, are very useful density diagnostics in   
a variety of astronomical sources. Though, computation of accurate transition rates, A-values, for these lines is difficult because the transitions are 
induced by mixing between the $^4$P and even parity doublet states. Hence,
the calculated A-values for the intercombination transitions depend critically on the quality of the representation of those even parity doublet states. 
Fig. \ref{intercomb} shows the evolution of the calculated $gf$-values in length 
and velocity forms from our calculations together with the results of
other authors and the experimental determinations of \cite{calamai}. 
Here we plot our results in the length and velocity gauges, although none of previous
authors present similar comparisons. It is clear, though, that there are considerable discrepancies 
between the two gauges, being the length form of the $gf$-values typically in better agreement
with experiment. Other authors, like \cite{tayal07}
do not even quote their results in the velocity gauge.
Indeed, the velocity form of the $gf$-value is typically less stable numerically,
since it depends on the first derivative of the radial wavefunction. Thus, we disregard the
velocity $gf$-values for subsequent analysis. The present $gf$-values from
{\sc autostructure} agree within 0.2~dex with 
experimental the determinations of \cite{calamai}, and the agreement between
MCDF results and experiment is even better, within 0.1~dex. On the other hand, the
HFR results seem as accurate as MCDF for some transitions, but significantly discrepant
for others. The results of \cite{tayal07} and \cite{froesefischer}, using a multiconfiguration Hartree-Fock, are
comparable in quality to our HFR results. The results of
\cite{dufton}, based on the multiconfiguration Hartree-Fock method, and \cite{nussbaumer},
from the use of the TFDA potential, seem to be of inferior quality.

It seems clear that no one calculation among those performed here and those
reported by other authors can provide ultimate accuracy $gf$-values for all
transitions simultaneously. Like with the different
calculations, there is significant scatter among the results of various experimental determinations. Thus, in order
to provide the most reliable set of $gf$-values possible we take the statistical
average among all theoretical results, i.e. our present results from AST8 and
AST9 and those of other authors. In computing the average we discard     
values that depart by more than 3$\sigma$ from the average, where $\sigma$
is the statistical dispersion of the

\begin{landscape}
\begin{table*}
\scriptsize
\caption[]{$\log(gf)$ values for transitions among the 15 lowest levels in \ion{Si}{ii}}      
\label{table3}
\centering
\begin{tabular}{llrccccccccc}
\hline\hline
Lower level & Upper level & $\lambda$ (\AA) & \multicolumn{2}{c}{AST8} & \multicolumn{2}{c}{AST9} &
HFR & \multicolumn{2}{c}{MCDF}& Recom. & Error \cr 
  &  &  & $\log(gf)_l$& $\log(gf)_v$ & $\log(gf)_l$ & $\log(gf)_v$ & $\log(gf)$ & $\log(gf)_C$& $\log(gf)_B$\cr
\hline
 3s$^2$3p $^2$P$^o_{1/2}$ & 3s3p$^2$  $^4$P$_{1/2}$&   2334.41& -5.00 & -4.89 & -4.89 & -4.55 & -4.98 &   -5.09 & -5.10 & -5.05 & 0.14 \cr
 3s$^2$3p $^2$P$^o_{3/2}$ & 3s3p$^2$  $^4$P$_{1/2}$&   2350.17& -5.08 & -4.66 & -4.94 & -4.86 & -5.13 &   -5.11 & -5.13 & -5.18 & 0.15 \cr
 3s$^2$3p $^2$P$^o_{1/2}$ & 3s3p$^2$  $^4$P$_{3/2}$&   2328.52& -7.72 & -7.28 & -7.49 & -8.06 & -7.19 &   -6.53 & -6.62 & -7.39 & 0.24 \cr
 3s$^2$3p $^2$P$^o_{3/2}$ & 3s3p$^2$  $^4$P$_{3/2}$&   2344.20& -5.32 & -4.42 & -5.27 & -4.59 & -5.15 &   -5.43 & -5.44 & -5.39 & 0.16 \cr
                          & 3s3p$^2$  $^4$P$_{5/2}$&   2334.20& -5.02 & -4.20 & -4.94 & -4.40 & -4.80 &   -4.81 & -4.85 & -4.99 & 0.20 \cr
 3s$^2$3p $^2$P$^o_{1/2}$ & 3s3p$^2$  $^2$D$_{3/2}$&   1808.00& -2.39 & -2.33 & -2.47 & -2.90 & -2.49 &   -1.99 & -2.03 & -2.31 & 0.16 \cr
 3s$^2$3p $^2$P$^o_{3/2}$ & 3s3p$^2$  $^2$D$_{3/2}$&   1817.45& -3.34 & -3.34 & -3.45 & -4.28 & -3.45 &   -2.82 & -2.87 & -3.14 & 0.26 \cr   
                          & 3s3p$^2$  $^2$D$_{5/2}$&   1816.92& -2.20 & -2.10 & -2.30 & -2.72 & -2.28 &   & & -2.17 & 0.22 \cr 
 3s$^2$3p $^2$P$^o_{1/2}$ & 3s$^2$4s  $^2$S$_{1/2}$&   1526.72& -0.24 & -0.38 & -0.46 & -0.51 & -0.60 &   & & -0.54 & 0.10 \cr 
 3s$^2$3p $^2$P$^o_{3/2}$ & 3s$^2$4s  $^2$S$_{1/2}$&   1533.45& +0.05 & -0.11 & -0.46 & -0.51 & -0.30 &   & & -0.27 & 0.09 \cr 
 3s$^2$3p $^2$P$^o_{1/2}$ & 3s3p$^2$  $^2$S$_{1/2}$&   1304.37& -2.62 & -2.07 & -0.71 & -1.01 & -0.64 &   & & -0.74 & 0.09 \cr 
 3s$^2$3p $^2$P$^o_{3/2}$ & 3s3p$^2$  $^2$S$_{1/2}$&   1309.27& -3.12 & -2.83 & -0.47 & -0.79 & -0.41 &   & & -0.49 & 0.09 \cr 
 3s$^2$3p $^2$P$^o_{1/2}$ & 3s$^2$3d  $^2$D$_{3/2}$&   1260.42& +0.39 & +0.32 & +0.49 & +0.40 & +0.38 &   & & +0.38 & 0.05 \cr 
 3s$^2$3p $^2$P$^o_{3/2}$ & 3s$^2$3d  $^2$D$_{3/2}$&   1265.02& -0.28 & -0.39 & -0.58 & -0.68 & -0.36 &   & & -0.40 & 0.10 \cr 
                          & 3s$^2$3d  $^2$D$_{5/2}$&   1264.73& +0.61 & +0.70 & +0.61 & +0.57 & +0.63 &   & & +0.62 & 0.04 \cr   
 3s$^2$3p $^2$P$^o_{1/2}$ & 3s3p$^2$  $^2$P$_{1/2}$&   1193.28& +0.041& -0.068& +0.060& +0.082& +0.069&   & & +0.037& 0.047\cr   
 3s$^2$3p $^2$P$^o_{3/2}$ & 3s3p$^2$  $^2$P$_{1/2}$&   1197.39& -0.24 & -0.22 & -0.34 & -0.31 & -0.21 &   & & -0.25 & 0.04 \cr 
 3s$^2$3p $^2$P$^o_{1/2}$ & 3s3p$^2$  $^2$P$_{3/2}$&   1190.42& -0.04 & -0.55 & -0.68 & -0.33 & -0.25 &   & & -0.29 & 0.10 \cr 
 3s$^2$3p $^2$P$^o_{3/2}$ & 3s3p$^2$  $^2$P$_{3/2}$&   1194.50& +0.39 & +0.52 & +0.39 & +0.38 & +0.48 &   & & +0.28 & 0.22 \cr   
 3s3p$^2$ $^4$P$_{1/2}$   & 3s$^2$4p $^2$P$^o_{1/2}$&  2605.62& -6.24 & -6.45 & -7.39 & -6.90 & -9.34 &   & & -6.56 & 0.34 \cr
 3s3p$^2$ $^4$P$_{1/2}$   & 3s$^2$4p $^2$P$^o_{3/2}$&  2601.56& -6.49 & -7.33 & -6.54 & -6.03 & -7.99 &   & & -6.50 & 0.45 \cr
 3s3p$^2$ $^4$P$_{3/2}$   & 3s$^2$4p $^2$P$^o_{1/2}$&  2613.00& -5.75 & -6.22 & -5.95 & -6.10 & -5.64 &   & & -5.88 & 0.21 \cr
 3s3p$^2$ $^4$P$_{3/2}$   & 3s$^2$4p $^2$P$^o_{3/2}$&  2608.91& -7.97 & -6.61 & -7.80 & -7.17 & -6.63 &   & & -6.94 & 0.40 \cr
 3s3p$^2$ $^4$P$_{5/2}$   & 3s$^2$4p $^2$P$^o_{3/2}$&  2620.90& -4.87 & -5.56 & -5.00 & -5.33 & -4.63 &   & & -4.97 & 0.30 \cr
 3s3p$^2$ $^2$D$_{3/2}$   & 3s$^2$4p $^2$P$^o_{1/2}$&  3862.60& -0.73 & -0.89 & -0.50 & -0.90 & -0.67 &   & & -0.86 & 0.28 \cr
 3s3p$^2$ $^2$D$_{3/2}$   & 3s$^2$4p $^2$P$^o_{3/2}$&  3853.66& -1.43 & -1.58 & -1.20 & -1.60 & -1.37 &   & & -1.55 & 0.28 \cr
 3s3p$^2$ $^2$D$_{5/2}$   & 3s$^2$4p $^2$P$^o_{3/2}$&  3856.02& -0.48 & -0.64 & -0.25 & -0.64 & -0.42 &   & & -0.62 & 0.32 \cr
 3s$^2$4s $^2$S$_{1/2}$   & 3s$^2$4p $^2$P$^o_{1/2}$&  6371.36& -0.37 & -0.99 & -0.11 & -0.21 & -0.12 &   & & -0.32 & 0.20 \cr
 3s$^2$4s $^2$S$_{1/2}$   & 3s$^2$4p $^2$P$^o_{3/2}$&  6347.10& -0.55 & -0.04 & -0.20 & -0.09 & +0.19 &   & & -0.01 & 0.19 \cr
\hline
\end{tabular}
\end{table*}
\end{landscape} 

\noindent{data.} Both, length and velocity $gf$-values,
are taken with equal weights in the average.
In the case of the
recombination transitions and the 1814\AA\ multiplet we also include
in
the averages the experimental data from \cite{calamai} and \cite{bergeson} respectively
with twice as much weight as the theoretical values.
It is noted that our calculation
AST8 that uses the smaller configuration expansion is very well optimized
on the lowest energy terms, for instance the 3s$^2$3p $^2$P$^o$ and
3s3p$^2$ $^2$D terms, but it deteriorates rapidly for higher excitation terms.
Fig. 2 depicts our recommended $gf$-values for every transition together
with the statistical dispersion, which is probably representative of the
true uncertainty.
In Table~3 we show 
our results for transitions among the lowest 15 levels of the ion 
from the AST9, AST9, HFR, and MCDF calculations. The
{\sc autostructure} results are given in length and velocity gauges and
the MCDF results are given in the corresponding Coulomb and
Babushkin gauges. In the last column of the table are
our recommended values and estimated uncertainties

\section{Collision Strengths}

The collision strengths for electron impact excitation are computed with
the ICFT Breit--Pauli $R$-matrix package ({\sc bprm}) based on the close-coupling
approximation of
\citet{bur71} whereby the wavefunctions for states
of an $N$-electron target and a colliding electron with total angular momentum
and parity $J\pi$ are expanded in terms of the target eigenfunctions
\begin{equation}\label{cc}
\Psi^{J\pi}={\cal A}\sum_i \chi_i{F_i(r)\over r}+\sum_jc_j\Phi_j\ .
\end{equation}
The functions $\chi_i$ are vector coupled products of the target eigenfunctions
and the angular components of the incident-electron functions, $F_i(r)$ are the
radial part of the projectile electron and $\mathcal{A}$ is an antisymmetrization operator.
The functions $\Phi_j$ are bound-type functions of the total system constructed
with target orbitals; they are introduced to compensate for orthogonality
conditions imposed on the $F_i(r)$ and to improve short-range correlations.
The Kohn variational principle gives rise to a set of coupled integro-differential
equations that are solved by $R$-matrix techniques
\citep{bur71,ber74,ber78,ber87} within a box of radius, say,
$r\leq a$. In the asymptotic region ($r>a$)
exchange between the outer
electron and the target ion can be neglected, and the wavefunctions can be
approximated by Coulomb solutions.

One-body Breit--Pauli relativistic corrections have been introduced in the
$R$-matrix suite by \citet{sco80,sco82}.
Inter-channel coupling is equivalent to CI
in the atomic structure context.

Because of the large number of configurations and close coupling states in the
representation of the target ion the scattering calculation had to be done
in LS-coupling. Then, fine structure collision strengths were obtained by
means of the intermediate coupling frame transformation (ICFT) method of
\cite{griffin}.

Since we have produced a large number of target expansion to study  
the quality and the wavefunction it is interesting to see
the effects of these various representations on the collision strengths.
This comparison may be used to asses the accuracy of the collision strengths.
Thus, we performed three different scattering calculations using target orbitals from models AST1, AST8, and AST9. For the first two calculations we included
only the lowest 12 LS terms in the closed coupling expansion, while for
the calculation with the AST9 target, the most accurate and extensive,
we included the lowest 43 LS terms. All calculations explicitly include
partial waves from states with $L\le 16$ and multiplicity 1, 3, 5, and 7.
The final collision strengths are produced with an energy resolution of
$6\times 10^{-5}$ Ry.
 
\begin{figure}
\resizebox{\hsize}{!}{\includegraphics{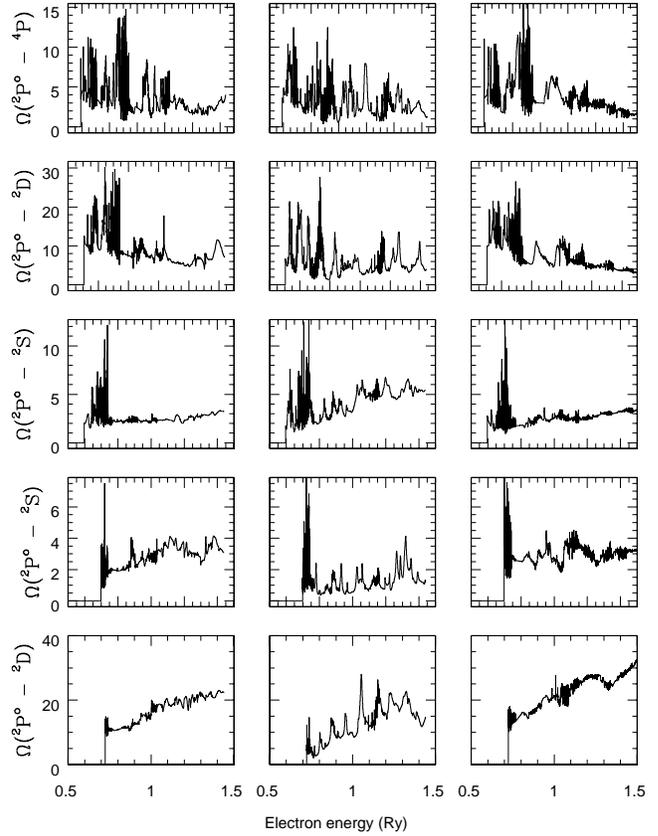}}
\caption{Comparison of collision strengths in LS coupling for excitations
from the ground term to the first five excite terms
in \ion{Si}{ii} computed with approximations AST1 (left panels),
AST8 (middle panels), and AST9 (right panels).} 
\label{omega1} 
\end{figure}

Figure \ref{omega1} compares the LS-coupling collision strengths obtained
from the target expansions AST1, AST8, and AST9 for
excitation from the ground term 3s$^2$3p $^2$P$^o$ to the first four
excited terms 3s3p$^2$ $^4$P, 3s3p$^2$ $^2$D, 3s$^2$4s $^2$S, and 
3s3p$^2$ $^2$S.

\begin{table}[h]
\centering
\caption[]{Comparison of Maxwellian averaged collision strengths in LS-coupling for
excitation from the ground term 3s$^2$3p~$^2$P$^o$.}  
\label{table4}
\begin{tabular}{lrrrr}
\hline\hline
Upper term & T (K) & Model 9& \multicolumn{2}{c}{Difference (\%)}\\
 & & & Model 1& Model 8\\
\hline
3s3p$^2$~$^4$P  &  5000 & 4.44E+0 &   1.2 &   5.7 \cr
                & 10000 & 4.36E+0 &   1.4 &   5.9 \cr
                & 20000 & 4.43E+0 &  -2.4 &  -5.2 \cr
3s3p$^2$~$^2$D  &  5000 & 1.19E+1 &  -8.2 &  -39.8 \cr
                & 10000 & 1.18E+1 &  -3.1 &  -36.4 \cr
                & 20000 & 1.10E+1 &   3.9 &  -34.2 \cr
3s$^2$4s~$^2$S  &  5000 & 2.17E+0 &  11.0 &   25.3 \cr
                & 10000 & 2.18E+0 &  16.3 &   27.0 \cr
                & 20000 & 2.21E+0 &  15.2 &   33.1 \cr
3s3p$^2$~$^2$S  &  5000 & 3.32E+0 & -38.1 &  -52.4 \cr
                & 10000 & 3.08E+0 & -32.7 &  -54.7 \cr
                & 20000 & 2.95E+0 & -24.4 &  -59.2 \cr
\hline
\end{tabular}
\end{table}

The results, in terms of Maxwellian averaged effective collision
strengths, of AST1, AST8, and AST9 are also compared in 
Table~4 for excitations
from the ground term, 3s$^2$3p~$^2$P$^o$, 
to the 
first five excited terms
of \ion{Si}{ii}. Here, the third column presents the effective collision strengths from our
best model AST9 and columns fourth and five present the percentage difference from these
as obtained from models AST1 and AST8 respectively. 
These comparisons are done for temperatures
between 5000 K and 20,000 K. Excitation rates for higher temperatures are
of little practical interest because under these conditions the ionic fraction in \ion{Si}{ii} is
too small.  
The are two general characteristics of collision strengths that could lead
to variations in their thermal averages: (i) the background and resonances
in the near threshold region, which determine the low temperature 
($\Delta E/kT\le 1$) Maxwellian averaged collision strengths and
depend on the coupling of the 
target ion with the continuum; and (ii) the slope of the continuum towards
high energies,
which in dipole allowed transitions depend linearly on the oscillator
strength and consequently on the quality of the target representation. 
It is interesting to see that the results from AST1 and AST9 agree rather well
in terms of the qualitative shape of the collision strengths and the  quantitative Maxwellian averaged rates,
within $\sim 10\%$, except for the 3s$^2$3p $^2$P$^o$ - 3s3p$^2$ $^2$S
transition. This is in contrast to the large differences in gf-values for essentially all transitions among levels of the $N$-electron \ion{Si}{ii} target.
The good agreement in collision strengths is because these are 
dominated by bound-continuum couplings of the $(N+1)$-electron system,
which seems to be well represented by the large amount CI in the
close coupling expansion

On the other hand, the target expansion AST8 does yield a different lay out  
of resonances on the collision strengths. In addition, AST8 yields steeper rises in the background collision strengths for excitation to the
3s$^2$4s $^2$S level and higher levels. This translates into effective collision
strengths from AST8 that are systematically lower than those from AST9
by as much as $\sim 50\%$.

It is expected that our results from the AST9 expansion 
should be the most accurate due to the good quality of the target
representation and the large amount of CI included for the 
(N+1)-electron system. Nonetheless, the comparison with results
of various target representation allow us to asses the uncertainties 
in the collision strengths to $\sim10\%$. 

\begin{figure} 
\resizebox{\hsize}{!}{\includegraphics{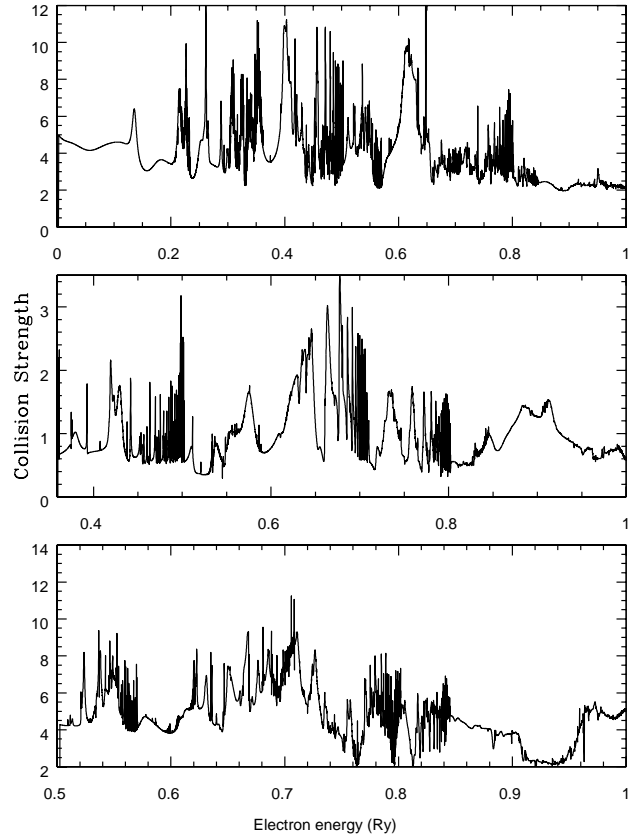}}
\caption{Collision strengths in JJ-coupling for transitions
3s$^2$3p~$^2$P$^o_{1/2}$~--~3s$^2$3p~$^2$P$^o_{3/2}$ (top panel),
3s$^2$3p~$^2$P$^o_{3/2}$~--~3s3p$^2$~$^4$P$_{3/2}$ (middle panel),
and 3s$^2$3p~$^2$P$^o_{3/2}$~--~3s3p$^2$~$^2$D$_{5/2}$ (bottomr panel).} 
\label{omega2}
\end{figure}

Fig.~\ref{omega2} shows the collision strengths for a sample of
forbidden, intercombination, and dipole transitions. We chose the same
transitions as in Tayal (2008) for direct comparison with that work.
Although the collision strengths from both calculations are roughly
similar, there are quantitative differences between the two sets in terms
of resonance structures and absolute level of the background.
It is apparent that the present collision strengths should yield 
Maxwellian averages somewhat lower than those of Tayal.

\begin{table}[h]
\centering
\caption[]{Comparison of Maxwellian averaged collision strengths in JJ-coupling 
from present calculation (Present), Tayal (2008; Tayal), and 
Dufton \& Kingston (1991; DK).}  
\label{table5}
\begin{tabular}{lrccc}
\hline\hline
Upper level & T (K) & Present & Tayal & DK\\
\hline
3s$^2$3p~$^2$P$^o_{3/2}$&  5000 & 4.55 &6.19&5.60 \cr
                        & 10000 & 4.45 &6.09&5.70 \cr
                        & 20000 & 4.42 &5.97&5.77 \cr
3s3p$^2$~$^4$P$_{1/2}$  &  5000 & 0.401 &0.512&0.550 \cr
                        & 10000 & 0.398 &0.515&0.516 \cr
                        & 20000 & 0.392 &0.502&0.466 \cr
3s3p$^2$~$^4$P$_{3/2}$  &  5000 & 0.612 &0.812&0.832 \cr
                        & 10000 & 0.609 &0.789&0.780 \cr
                        & 20000 & 0.602 &0.769&0.706 \cr
3s3p$^2$~$^4$P$_{5/2}$  &  5000 & 0.441 &0.615&0.571 \cr
                        & 10000 & 0.458 &0.595&0.534 \cr
                        & 20000 & 0.477 &0.589&0.488 \cr
3s3p$^2$~$^2$D$_{1/2}$  &  5000 & 1.82 &2.77&2.76 \cr
                        & 10000 & 1.82 &2.74&2.74 \cr
                        & 20000 & 1.75 &2.50&2.58 \cr
3s3p$^2$~$^2$D$_{3/2}$  &  5000 & 2.05 &2.94&2.45 \cr
                        & 10000 & 2.14 &2.98&2.44 \cr
                        & 20000 & 2.14 &2.80&2.30 \cr
3s$^2$4s~$^2$S$_{1/2}$  &  5000 & 0.910 &1.02&1.24 \cr
                        & 10000 & 0.865 &1.06&1.20 \cr
                        & 20000 & 0.857 &0.979&1.04 \cr
3s3p$^2$~$^2$S$_{1/2}$  &  5000 & 0.887 &0.102&0.716 \cr
                        & 10000 & 0.899 &0.988&0.840 \cr
                        & 20000 & 0.916 &0.988&0.902 \cr
\hline
\end{tabular}
\end{table}

In Table~5 we compare the present fine structure Maxwellian
averaged collision strengths with those of Tayal (2008) and Dufton
\& Kingston (1991). For most transitions our results are $\sim 30\%$
lower than those of Tayal, while the results of Dufton \& Kingston lie
inbetween those two. That our results are somewhat lower than
those of Dufton \& Kingston can be understood from the much larger
close coupling expansion used in the present work. As more scattering
channels are open in the calculation, the flow of electrons is
redistributed and the collision strength among low lying levels tends
to converge to lower values. The source of the differences found with respect to
Tayal are less clear, and seems to be due to differences in computational approach
used, i.e. between the orthogonal and non-orthogonal R-matrix methods.

\section{Conclusions}

We have carried out extensive calculations of transitions rates and
collision strengths for electron impact excitation for the lowest 
12 levels of the astrophysically important \ion{Si}{ii} ion.

In the calculation of radiative data,
we paid special attention to the weak dipole allowed and intercombination
transitions, which are of particular interest for plasma diagnostics.
Determination of accurate data for these transitions is particularly
challenging, therefore we studied the effects of valence-valence,
valence-core, and core-core interactions
with three different methods, i.e. MCDF, HFR, and the central potential
method implemented in {\sc autostructure}.
With MCDF we could only include valence-valence correlations, as
opening of the n=2 core resulted in a large number of states that
could not be managed with the computer code {\sc grasp}. For this reason,
we were able to obtain accurate transition rates for the 
3s$^2$3p~$^2$P$^o$~--~3s3p$^2$~$^4$P intercombination transitions only.
Both HFR and {\sc autostructure} allowed us to investigate valence-core
and core-core interaction by building very large configuration expansions.
The accuracy of {\sc autostructure} calculations were significantly improved 
by the use of the modified TFDA potential of \cite{bautista08} and a
new technique for optimizing the variational parameters of this potential.
This optimization technique takes into account the differences between 
length and velocity gauges of the $gf$-values, in addition to the accuracy
of predicted energy levels.
Our most accurate $gf$-values were then compared with previous theoretical and
experimental determinations. From these comparisons we derive a recommended 
set of $gf$-values and estimate their uncertainties.

We then proceed to compute electron impact excitation collision strengths
with the R-matrix method. We do so by using various of the targets
representations made from the previous calculations and compare the results.
This allows us to identify the physical effects that affect the
accuracy of the computed collision strengths. We also compare
our results with those of previous calculations.
The present results agree reasonably well with those of Dufton \&
Kingston (1991), who also used an orthogonal R-matrix method but with a much 
smaller close coupling expansion. On the other hand, the present 
results for Maxwellian averaged collision strengths are systematically
lower than those of the recent calculation of Tayal (2008) using
a non-orthogonal R-matrix method. The reasons for these differences,
that typically amount to $~\sim 30\%$, are unclear. We argue that
the source of this difference
could be in the non-orthogonal R-matrix approach.

Tables 6 and 7 containing the present gf-values, A-values, and effective collision strengths are available in electronic form. The data and atomic model are also to become available through the 
TIPTOP\footnote{\tt http://heasarc.gsfc.nasa.gov/topbase} database
and the XSTAR database \cite{bautistakallman01}.

\begin{acknowledgements}
MAB acknowledges financial support from grants
from the NASA Astronomy and Physics Research and Analysis Program (award Award NNX09AB99G ) and
the Space Telescope Science Institute (project GO-11745).
Financial support from the Belgian F.R.S.-FNRS is also acknowledged by two of us (PQ and PP) who are, respectively, Senior Research Associate and Research Associate of this organization. 
\end{acknowledgements}


\end{document}